\PassOptionsToPackage{dvipsnames}{xcolor}
\documentclass[conference]{IEEEtran}

\usepackage{xspace}
\usepackage{listings}
\usepackage{subcaption}
\usepackage{todonotes}
\usepackage{url}
\usepackage{tcolorbox}
\newcommand{\jess}{\textsc{Jess}\xspace}

\newcommand{\llbox}[1]{
		\begin{tcolorbox}[width=\columnwidth, colframe=black, boxrule=0.25mm, top=1mm, left=1mm, right=1mm, bottom=1mm]
			#1
		\end{tcolorbox}
}

\newif\ifshowcomments

\begin{document}

\title{Compilation of Commit Changes within Java Source Code Repositories}

\author{
	\IEEEauthorblockN{Stefan Schott}
	\IEEEauthorblockA{\textit{Paderborn University}\\
		Paderborn, Germany \\
		stefan.schott@upb.de}
	\and
	\IEEEauthorblockN{Wolfram Fischer}
		\IEEEauthorblockA{\textit{SAP Security Research}\\
			Mougins, France \\
			wolfram.fischer@sap.com}
	\and
	\IEEEauthorblockN{Serena Elisa Ponta}
	\IEEEauthorblockA{\textit{SAP Security Research}\\
		Mougins, France \\
		serena.ponta@sap.com}
	\and
	\IEEEauthorblockN{Jonas Klauke}
	\IEEEauthorblockA{\textit{Paderborn University}\\
		Paderborn, Germany \\
		jonas.klauke@upb.de}
	\and
	\IEEEauthorblockN{Eric Bodden}
	\IEEEauthorblockA{\textit{Paderborn University \& Fraunhofer IEM}\\
		Paderborn, Germany \\
		eric.bodden@upb.de}
}

\maketitle

\begin{abstract}
	Java applications include third-party dependencies as bytecode.
To keep these applications secure, researchers have proposed tools to re-identify dependencies that contain known vulnerabilities.
Yet, to allow such re-identification, one must obtain, for each vulnerability patch, the bytecode fixing the respective vulnerability at first.
Such patches for dependencies are curated in databases in the form of fix-commits.
But fix-commits are in source code, and automatically compiling whole Java projects to bytecode is notoriously hard, particularly for non-current versions of the code.

In this paper, we thus propose \jess, an approach that largely avoids this problem by compiling solely the relevant code that was modified within a given commit.
\jess reduces the code, retaining only those parts that the committed change references.
To avoid name-resolution errors, \jess automatically infers stubs for references to entities that are unavailable to the compiler.
A challenge is here that, to facilitate the above mentioned re-identification, \jess must seek to produce bytecode that is almost identical to the bytecode which one would obtain by a successful compilation of the full project.

An evaluation on 347 GitHub projects shows that \jess is able to compile, in isolation, 72\% of methods and constructors, of which 89\% have bytecode equal to the original one.
Furthermore, on the \emph{Project KB} database of fix-commits, in which only 8\% of files modified within the commits can be compiled with the provided build scripts, \jess is able to compile 73\% of all files that these commits modify.

\end{abstract}


	\section{Introduction}
\label{sec:introduction}

Java projects on average comprise 71\% of code from open-source libraries~\cite{stateOfDependencyManagement2023}, many of which contain known, exploitable vulnerabilities.
Under the name of software composition analysis, commercially available tools thus try to re-identify such known-to-be-vulnerable code when it gets included into a given Java project.
Yet, as Dann et al.\ showed~\cite{dann2021identifying}, detection is commonly performed by checking only the metadata associated with the included dependency.
This metadata is then compared against a database, which contains information about vulnerable versions of certain dependencies.
However, as Ponta et al.~\cite{ponta2018beyond} have shown, because metadata is often unavailable or inaccurate, these tools frequently perform a misclassification.
To improve over this, one must seek to unambiguously re-identify the actually \textit{vulnerable code} within the application's dependencies.
Existing vulnerability databases such as Project KB~\cite{ponta2019manually} already map known vulnerabilities to commits in the related source code repositories.
These existing vulnerability databases contain the exact source code repository \textit{snapshot} of the applied fix.
But Java projects typically include dependencies as bytecode, not source code.
Further, as Dann et al.~\cite{dann2021identifying} showed, many open-source dependencies are rebundled or repackaged, i.e., their JAR files contain bytecode they do not declare to contain.
To unambiguously determine whether a given piece of vulnerable code does or does not exist in a given bytecode-level dependency, one thus must be able to reason about this vulnerability in terms of its bytecode.
To this end few approaches like BScout~\cite{dai2020bscout} and PPT4J~\cite{pan2024ppt4j} have been developed that try to address this problem by trying to compare the included bytecode to the source code of the respective patches.
However, they acknowledge this source-to-bytecode comparison to be specifically challenging, which causes imprecisions in the detection.

Now one might think it to be easy to just compile the history version of the project that implemented the given vulnerability's patch to avoid the source-to-bytecode comparison.
But unfortunately it is notoriously hard to compile Java projects obtained from source code repositories~\cite{hassan2017automatic, sulir2016quantitative, tufano2017there}.
As mentioned above, existing vulnerability databases only contain the repository snapshot, but not the release version, where the patch is applied, and being able to obtain the release version corresponding to a snapshot is highly dependent on the employed development practices within the repository (e.g. use of proper tags, semantic versioning, etc.). 
Therefore one needs to be able to directly compile the snapshot representing the vulnerability patch.
However, compiling a historic snapshot from a source code repository is one particularly hard problem.
As Tufano et al.~\cite{tufano2017there} have shown, across 100 investigated Java projects from the Apache Software Foundation, which is known for an often outstanding maintenance of projects, only 38\% of the change history was compilable.

Yet, luckily, when the task is to re-identify known-to-be-vulnerable code, one does not require the full code of the snapshot, but only the code of the security-relevant methods that a given fix-commit actually patches~\cite{dai2020bscout, pan2024ppt4j, ponta2018beyond}.
To this end, we propose \jess, an approach that reduces the code to be compiled by retaining only the code that a given fix-commit modifies.
\jess receives source code of a Java project as input and slices away all information from the code base that is not required to compile the specific areas of interest.
If done na\"ively, this can easily cause the compiler to yield name-resolution errors.
To avoid these, \jess infers the types that are required for the compilation to succeed, yet which are not available within the code base, and generates type stubs that correspond to these inferred types.
A challenge is here that, to facilitate the re-identification of vulnerabilities in the resulting bytecode, \jess must seek to produce bytecode that is not just similar but actually identical to that which one would obtain also by a successful compilation of the full project.
Using \jess one can easily obtain the bytecode corresponding to a vulnerability patch, which can subsequently be used by software composition analysis approaches without the need to employ a challenging source-to-bytecode comparison.

We evaluated \jess on 347 of the most popular Java projects on GitHub.
Our evaluation showed that of the 32,970 methods and constructors that we had randomly sampled from the projects, \jess was able to compile more than 72\% in isolation.
The evaluation further showed that the bytecode obtained via \jess is highly similar to the original bytecode.
It only differs by 0.66\% on average, and in 89\% of the cases it even equals the bytecode obtained by full compilation.
Further, our evaluation on a real-world dataset of fix-commits~\cite{ponta2019manually}, in which only 8\% of files modified in the commits can be compiled with the provided build scripts, shows that \jess is nonetheless even here able to compile 73\% of all files that these commits modify.

To summarize, the original contributions of this paper are:
\begin{itemize}
	\item An approach to compiling areas of Java code, modified within a commit, based on slicing, type inference and stubbing, with the goal to produce bytecode identical to that of the original compilation.
	\item An implementation in an open-source tool\footnote{\url{https://github.com/stschott/jess}}\footnote{\url{https://doi.org/10.5281/zenodo.12804723}}.
	\item A large-scale evaluation on 347 of the most popular Java projects on GitHub and the Project KB~\cite{ponta2019manually} fix-commit dataset.
\end{itemize}
	\section{Compiling Commit Changes with \jess}
\label{sec:concept}

\begin{figure}
	\includegraphics[width=\linewidth]{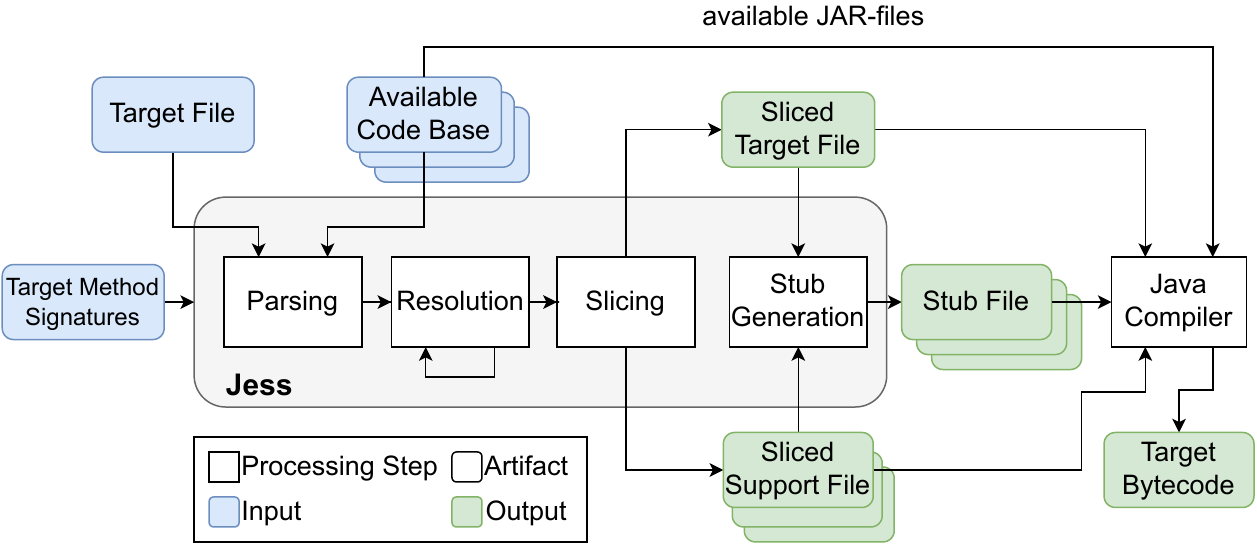}
	\caption{Overview of \jess}
	\label{fig:overview}
\end{figure}

Figure~\ref{fig:overview} shows our approach behind compiling commit changes with \jess. 
The goal is to obtain bytecode identical or at least very close to the one that would be generated by invoking the original build scripts with every single dependency fully resolved.
\jess takes as input the signatures of the methods to be compiled (referred to as target in the following), the Java source code files containing them, all other Java source code files available in the hosting code base and all dependencies that are still retrievable in form of JAR-files.
For the sake of brevity we describe the approach using a single method as target, however \jess supports handling of single or multiple members (e.g., methods, fields, constructors, and initializers) of classes and inner classes.

\jess follows a mark-and-sweep approach~\cite{mccarthy1960recursive} for its processing.
Initially, \jess parses all available files and marks the target with the annotation \textit{Keep-All} to denote that it must not be altered (e.g., the body of methods must not be sliced or modified).

In the resolution step, \jess identifies the parts of the code base that are relevant for compiling the target. It inspects all references in the target, e.g., field accesses, method invocations, or object creations, and tries to find the corresponding declaration. Such references may be part of the target file itself, a different file within the code base, or external files. Referenced members are marked with the annotation \textit{Keep} to denote that only the signature of the member should be kept.
All newly marked members are recursively resolved until no further references are found or the found references are not part of the available code base.
While references in the signatures are always resolved, those in the body of methods and constructors only need to be resolved when annotated as \textit{Keep-All}. For fields, the type definition is always resolved, whereas the field initializer is resolved only in case of \textit{Keep-All}.

In the slicing step, \jess removes all class members that are not needed for the compilation of the target: 
class members that have not been marked, bodies of methods and constructors marked as \textit{Keep} only, and the initializers of fields marked as \textit{Keep} only.
Additionally, \jess removes classes and inner classes without any marked member, now unused import declarations, and JavaDoc as it is not reflected within the bytecode.
Exceptions that need to be considered while slicing are described in Section~\ref{sec:slicing}.
Slicing the body of methods invoked from the target method, while keeping its signature, allows to minimize the amount of transitively referenced members, while obtaining compilable code.
After slicing, \jess outputs the sliced target file, which contains the target method and the class members it references, and a set of sliced support files, which contain class members that are either referenced directly or transitively from the target method.

Due to slicing away code unrelated to the target methods compilation, it may enable a successful compilation of the target method, even when the snapshot contains broken source code (e.g. due to syntax errors), as long as the the broken code is not contained within the target method itself.
After slicing, the compilation may still fail due to references to classes outside of the available code base.

Finally, the stub generation step infers the signature of unresolved references based on the symbol's usage context and synthesizes a \textit{type stub}, i.e., definitions of classes and class members with full signatures, yet no content. 
Similar to interfaces, which describe the structure of a class without describing the specific implementations, \jess generates classes that contain the full member signatures without any actual functional implementation. 
The stub generation is described in detail in Section~\ref{sec:stubGeneration}.
Note that after applying the slicing and stub generation steps, even though it might affect the resulting bytecode of other classes and methods that are not declared as target, \jess does not perform another backward impact analysis.
Current software composition analysis approaches perform a syntactic matching of only the fix code~\cite{dai2020bscout, pan2024ppt4j, ponta2018beyond}. 
Thus, \jess only aims at producing (equal) bytecode for the member declared as target, while the dependent files are only used to enable a compilation.

Finally, a Java compiler can then compile the target method with the sliced target file, sliced support files, generated stub files and available JAR-files.

\section{\jess Compilation Example}
\label{sec:example}

\renewcommand{\figurename}{Listing}
\noindent
\begin{figure}
	\begin{minipage}[]{.45\linewidth}
		\begin{lstlisting}
import org.example.X;

public class A {
  @Keep // 2
  int num = D.getNumber();

  int otherNum = 12;

  @KeepAll // 1
  public int visit() {
    B b = new B();
    b.inc(num);

    X x = new X();
    String s = "five";
    int val = x.getVal(s);
    boolean con = X.conv;

    return val;
  }
}
		\end{lstlisting}
		\subcaption{Target file}
		\label{lst:classA}
	\end{minipage}
	\hfill
	\begin{minipage}[]{.45\linewidth}
		\begin{lstlisting}
public class B {
  @Keep // 2
  public C inc(int a){...}
  public E dec(int a){...}
  
  @Keep // 3
  class C {...}
}
		\end{lstlisting}
		\subcaption{Support file}
		\label{lst:classB}
		\begin{lstlisting}
// Generated Stub
package org.example;

public class X {
  static boolean conv = true;

  public int getVal
    (String arg) {return 0;}
}
\end{lstlisting}
\subcaption{Generated stub file}
\label{lst:classX}
	\end{minipage}
	\caption{\jess compilation example}
	\label{lst:tarCompExample}
\end{figure}
\renewcommand{\figurename}{Fig.}

Listings~\ref{lst:classA},~\ref{lst:classB} and~\ref{lst:classX} illustrate an example of how \jess performs the processing for compiling commit changes.
Listing~\ref{lst:classA} shows the target file with the \texttt{visit} method being the target method (i.e. modified within the commit).
Only the classes \texttt{A} and \texttt{B} (cf. Listings~\ref{lst:classA},~\ref{lst:classB}) are assumed to be within the available code base.
After parsing, in the first iteration of the resolution step, \jess marks the target method as \textit{Keep-All} (e.g. method changed in the commit) to state that the method body should be kept intact.
The numbers next to the markings within the listings indicate the resolution iteration in which the corresponding member has been marked.
In the next iteration, \jess resolves all references within the \texttt{visit} method.
In line~11 (Listing~\ref{lst:classA}), a new object of class \texttt{B} is created and in line~12, the \texttt{inc(int a)} method of class \texttt{B} is invoked.
Due to this invocation, \jess marks the method declaration (line~3 in Listing~\ref{lst:classB}) as \textit{Keep}, which indicates that the method signature needs to be kept intact, while the body can be sliced away.
This method invocation receives the field \texttt{num} as argument and therefore \jess marks the field declaration in line~5 of Listing~\ref{lst:classA} also with a Keep-Marking.
The remaining references within the target method's body or signature all reference a class \texttt{X}, which is assumed to not be part of the code base.
Now \jess looks at all members that have been marked in the previous step.
As all added markings are Keep-Markings, only the member signatures are of interest, since their content will be sliced away in the next step.
Thus, the only additional reference is in line~3 of Listing~\ref{lst:classB} where
the \texttt{inc} method specifies a return type of \texttt{C}.
Because of this, a Keep-Marking is added to the inner class C (cf. line~6 of Listing~\ref{lst:classB}).
After all relevant references for the compilation of the target method have been considered, \jess removes all non-marked members within the classes and the contents of the members marked with a Keep-Marking.
In class \texttt{A} \jess will remove the declaration of the field \texttt{otherNum} and the initialization of the field \texttt{num}.
In class \texttt{B} \jess will remove the declarations of the method \texttt{dec}, as well as the body of method \texttt{inc} (and add a corresponding dummy return statement, which will be explained in detail in Section~\ref{sec:slicing}).
Finally, in inner class \texttt{C}, \jess will remove all class members.
In doing so, \jess eliminates references to the classes \texttt{D} and \texttt{E}, which might be originating from an external library that might not be available any longer and would cause a normal compilation of the full snapshot to fail.

After slicing, not all references to classes outside of the available code base could be removed.
As the target method  contains multiple references to members of the unavailable class \texttt{X}, 
\jess now generates a stub file, which can be seen in Listing~\ref{lst:classX}.
When \jess processes the creation of the non resolvable object (line~14 in Listing~\ref{lst:classA}), it tries to infer all necessary information to generate an appropriate stub file from its usage context.
In this case,  since no argument is given in the object creation, \jess does not need to generate any constructor.
To generate the correct bytecode, however, the simple name of the referenced class is insufficient---the fully qualified name (FQN) is required instead.
To determine the FQN, \jess considers the import declaration within the target file (line~1 in Listing~\ref{lst:classA}) and sets the package definition of the generated stub file to the determined package path (\texttt{org.example} in this example).
The next unresolvable reference can be seen in line~16 of Listing~\ref{lst:classA}, where the method \texttt{getVal} is invoked.
Whenever \jess processes a method invocation, it needs to determine four properties:
\begin{enumerate}
	\item the object or class that the method is being invoked on,
	\item the return type of the method,
	\item the parameter types of the method,
	\item whether the method is static or non-static.
\end{enumerate}
In the usage context within Listing~\ref{lst:classA}, it is possible to determine all four properties.
The method is invoked on an object with the name \texttt{x}, which can be traced back to the object instantiation in line~14.
The return value of the method is assigned to a variable of type \texttt{int}.
As its only argument it receives the variable \texttt{s}, which can be traced back to its definition site in line~15, where it is defined as type \texttt{String}.
Since the method is invoked on the previously instantiated object \texttt{x}, it is likely that the method is non-static.
Due to this information, \jess generates the corresponding method stub in lines~7--8 in Listing~\ref{lst:classX}.
Due to syntactic ambiguities (which will be described in detail in Section~\ref{subsec:ambiguities}) it is not possible to infer with certainty the definite return and parameter types and whether the method is static or non-static.
However, the generated stub will still enable a successful compilation.
Finally, the reference within the target method can be observed in line~17 in Listing~\ref{lst:classA}.
Here a field of class \texttt{X} is accessed and assigned to the \texttt{boolean} variable \texttt{con}.
Due to the usage context, \jess is able to generate the field stub in line~5 of Listing~\ref{lst:classX}, by applying the same inference procedure as for the previous method invocation.
When the sliced target file, sliced support files, and the generated stub file are handed to a Java compiler, the target method can be successfully compiled.

\section{Slicing Exceptions}
\label{sec:slicing}

In the following we describe some exceptions to be considered during the slicing step.

	\textbf{Abstract interface / superclass methods:} Whenever a class implements an interface or extends an abstract class, every abstract method definition needs to be implemented in concrete subtypes, otherwise the compilation will fail. 
	Because of this, \jess can only slice away methods within classes that are implementing an abstract method if it also slices away the corresponding abstract method definition from within the interface or superclass.
	However, if the interface or superclass is originating from an already-compiled class file or the JDK, e.g. the \texttt{Collection} interface, \jess cannot slice away any such method definitions.
	In such cases, \jess only slices away the body of such methods.

	\textbf{Abstract functional interface methods:} A functional interface in Java is an interface that defines just one abstract method~\cite{interfaces}.
	Whenever this is the case, a lambda or method reference expression can be used to create instances of the corresponding interface.
	In such cases, although the abstract method is typically not referenced explicitly, its definition still needs to be kept such that its defining interface is considered as functional by the Java compiler.
	Thus, \jess does not slice away the abstract method from within referenced functional interfaces, even when the method is not referenced explicitly.
	If the interface is not referenced at all, \jess will slice away the whole interface.

There are also exceptions that apply to slicing the bodies and initializers of methods, constructors and fields:

	\textbf{Method return types:} Whenever a method signature specifies a return type that is not \texttt{void}, the corresponding method body needs to contain a return statement, which returns a value that adheres to the specified type.
	When this is the case, \jess removes the whole method body and replaces it with a single return statement that returns a dummy-value (e.g. \texttt{return 0} or \texttt{return ``dummy''}).
	If the specified return type is a reference type (except for String types), \jess inserts a \texttt{return null} statement.
	
	\textbf{Super constructor invocation:} Whenever a class extends another class, the Java compiler requires the extending class to (transitively) call a constructor of the super class (via the \texttt{super} statement)~\cite{super}.
	If the super class contains a default constructor (which does not require parameters) this call is automatically generated by the Java compiler.
	However, when the super class only provides constructors that expect arguments, a super constructor needs to be called explicitly from within a constructor belonging to the extending class.
	Thus, \jess keeps such super constructor invocations intact and only replaces their arguments with dummy-values. 
	
	\textbf{Final fields:} Whenever a field is declared as \texttt{final}, it needs to be assigned a value during instance initialization.
	Thus slicing away the initialization of a final field would result in the Java compiler not being able to compile the respective class.
	Furthermore, final fields, which are initialized with a literal or an arithmetic expression (e.g. \texttt{final int x = y + 1}), are treated as constants by the Java compiler, with the constant being propagated directly to the method referencing it.
	Therefore, it is not sufficient to replace the initialization with a dummy value, as this will modify the resulting bytecode.
	Due to this, if the field is initialized with a literal or arithmetic expression, \jess keeps the original initialization intact.
	
	\textbf{Type-import-on-demand:} Whenever there is a type-import-on-demand declaration (import with asterisk), \jess keeps the import declaration as it might contain important information for the stub generation step.
	After stub generation, to avoid a compilation fail due to the package path not existing, \jess generates the directory structures described by these import declarations.

\section{Stub Generation}
\label{sec:stubGeneration}

In the following we describe the generation of type stubs, characterized by a full signature and no content.

For generating fitting type stubs, \jess applies a greedy algorithm.
First, it scans the code parts that remain after slicing for references that could not be resolved within the available code base.
For each unresolvable reference, based on the usage context, \jess then applies a set of inference techniques to determine corresponding types.
The determined types are required to create type stubs, which will not only satisfy the Java compiler, but result in bytecode that is as close as possible to the original one.
Finally, once all unresolvable references have been considered and the appropriate type information has been inferred, \jess generates the respective stub files containing the type stubs.
\jess applies the following techniques to infer appropriate stubs for references which are not resolvable within the available code base:

	\textbf{Literal inference:} The easiest case for \jess to infer the needed types for, e.g.,\ a method signature, is if the method has been invoked with a literal.
	The Java language specification clearly describes the types of literals~\cite{expressions}.
	For a method invocation, e.g., \texttt{foo("abc")}, which is invoked with the literal \texttt{"abc"}, the Java specification defines the type of the literal to be \texttt{String}.
	
	\textbf{Definition inference:} During definition inference, \jess tries to locate the definition-site of a variable and infers its type from there.
	This definition-site can either be a variable, field or parameter definition.
	Here it is important to consider the scopes of variables.
	At first \jess checks the block in which the variable has been used and then incrementally checks outer blocks, where the scope still applies to the variable usage, until it reaches the method definition.
	At this point \jess checks the method parameters.
	If the definition-site still has not been found, \jess incrementally increases its scanning towards field definitions within the same class, potential outer classes and extended superclasses.
	
	\textbf{Conditional inference:} Whenever an expression is used in a statement that requires a condition, e.g. if-, while- or do-statement, it has to resolve to a boolean type~\cite{blocks}.
	This means that if e.g. a method's return type is used as the condition of an if-statement, its return type has to be \texttt{boolean} or \texttt{Boolean}.
	
	\textbf{Return statement inference:} When an expression is used in a return statement, the type of the expression has to be the same (or more specific) as the return type defined in the method signature.
	
	\textbf{Static / Non-static inference:} To check if a method or field declaration is static, \jess checks if it finds a definition-site for the object/class the method or field is accessed on.
	If it finds a definition-site, the field/method is declared as non-static, otherwise as static.
	
	\textbf{FQN inference:} To infer the fully qualified name (FQN) of a class, \jess checks all direct import statements within a file to find one where the last part of the import matches the used class.
	Sometimes classes are imported via asterisk imports.
	If none of the direct import statements matched and there is only a single asterisk import statement in the file, \jess infers the FQN from the asterisk import.
	
	\textbf{Superclass / Interface inference:} There are some scenarios in which the compilation will fail if a certain class is not extended or a certain interface is not implemented.
	\begin{itemize}
		\item \textbf{Exceptions:} When objects are used in throw-statements or caught in try-catch blocks, the Java compiler requires the corresponding classes to extend the respective \texttt{Throwable} class.
		If \jess observes such usages, it adds the corresponding class extension to the generated stubs.
		\item \textbf{Iterable:} If an object is used in a for-each loop, the Java compiler requires the corresponding class to implement the \texttt{Iterable} interface.
		\jess detects such occurrences and adds the appropriate \texttt{Iterable} interface implementation.
		\item \textbf{AutoCloseable:} When an object is used as resource within a try-with-resources statement, the Java compiler requires the corresponding class to implement the \texttt{Auto\-Closeable} interface.
		\jess adds such an interface implementation to the generated stub class if needed.
		\item \textbf{Typecasts:} Whenever an explicit typecast is performed on an object, its corresponding class must inherit or implement the type defined within the typecast.
		\jess either adds the corresponding interface implementation or the respective extend class statement to the generated class.
	\end{itemize}

	\textbf{Generics inference:} Sometimes, when objects are created, generic type arguments are supplied to the created object.
	However, in order to be able to pass generic type arguments, the respective class needs to specify corresponding generic type parameters.
	When \jess observes such occurrences, it generates the stub class in a way, that allows for supplying generic type arguments to it.

\subsection{Type Inference Ambiguities}
\label{subsec:ambiguities}

A precise inference of the corresponding signature is not always possible, due to ambiguities that arise if some information is missing.
Dagenais and Hendren~\cite{dagenais2008enabling} already described some of the ambiguities that arise in the syntax of partial Java programs.
They already describe that precisely inferring the FQN is not possible, when there is no direct import of a class, but multiple asterisk imports are present within the target file.
Furthermore, they describe that e.g. it is not always possible to distinguish package names from class names (in case of inner classes).
However, the list presented is not exhaustive as we found other ambiguous syntax constructs: 

\textbf{Method invocations without return value usage:} When a method is invoked, but its return value is not used, it is unclear what return type has been originally specified in the method declaration.
While it is likely that the method does not return any value (i.e. \texttt{void} return type), it is still possible that a return type is specified, but the return value is simply dropped after its invocation.
Listing~\ref{lst:amb1} shows an example of such an occurrence.
The return value of the \texttt{find} method, which is invoked in line~4, is not used.
However, the possible method stubs in line~10 and line~12 are both valid, even though one returns a value of type \texttt{String}, while the other does not.
In each scenario the Java compiler generates different bytecode.

\textbf{Method invocations with literal arguments:} When a method is invoked with a literal as argument it is often not possible to unambiguously infer the specified parameter type within the method declaration.
If a method is invoked with the integer literal \texttt{5} as argument, the original method declaration might e.g. specify \texttt{int}, \texttt{long}, \texttt{float} or \texttt{double} as parameter type.
If a method is invoked with a \texttt{null} literal, it can only be inferred that the specified type is not primitive.
One example of this can be again observed in Listing~\ref{lst:amb1}, as the invoked \texttt{find} method can either specify \texttt{int} or \texttt{long} as parameter type and still be valid.

\textbf{Static vs. non-static:} As static methods or fields can either be accessed on a class itself or object instances of a class, it is not possible to precisely distinguish static class members.
While it is likely that members accessed on an object instance are non-static, there is no guarantee.
An example of this can be seen in Listing~\ref{lst:amb1}.
It is unclear whether the in line~4 invoked \texttt{find} method is declared as static or non-static, since both possibilities are valid.

\textbf{Generic class methods:} In some scenarios, when a method belonging to a generic class is invoked, it is unclear whether the method specifies its parameters explicitly or as a generic type.
Listing~\ref{lst:amb2} illustrates such an example.
In line~3 an object of type \texttt{W} is created with \texttt{String} supplied as the generic type argument.
Then in line~4 the method \texttt{add} is invoked on the generic object, with a string literal as argument.
Now, if we look at the stub, it is possible that the method \texttt{add} specifies a generic type parameter (see line~10) or explicitly specifies \texttt{String} as its parameter type (see lines~12 \& 13).
Each scenario results in different bytecode being generated.

\textbf{Interface vs. Class:} Without having additional context, it is not always possible to distinguish whether a specified type is originating from a class or an interface.
When a method is called on an object of such a type, in the bytecode it is either invoked via an \texttt{invokevirtual} instruction, in case of originating from a class, or via an \texttt{invokeinterface} instruction, in case of originating from an interface.

\textbf{No explicit return type usage:} It may be the case that e.g. a method outside of the available code base is invoked and its return value is assigned to a variable that is declared using the \texttt{var} keyword, which does not explicitly assign a type to the variable.
In this case is is not possible to precisely determine the return type of the respective method.
The same applies to nested method invocations where two methods are outside of the available code base and the return value of one method is directly used as argument for the other method, without being assigned to a variable beforehand (e.g. \lstinline|foo(bar())|).
In this case it is not possible to precisely infer the return type of one method and the specific parameter type of the other method.

\renewcommand{\figurename}{Listing}
\noindent
\begin{figure}
	\begin{minipage}[t]{.45\linewidth}
		\begin{lstlisting}
public class G {
  void process() {
    V v = new V();
    v.find(5);
  }
}

// Stub
public class V {
  public void find(int i) {}
  // or
  public static String find
    (long i) {return "abc";}
}
		\end{lstlisting}
		\subcaption{Ambiguous example 1}
		\label{lst:amb1}
	\end{minipage}
	\hfill
	\begin{minipage}[t]{.45\linewidth}
		\begin{lstlisting}
public class H {
  void verify() {
    W<String> w = new W<>();
    w.add("value");
  }
}

// Stub
public class W<T> {
  public void add(T t) {}
  // or
  public void add
    (String s) {}
}
		\end{lstlisting}
		\subcaption{Ambiguous example 2}
		\label{lst:amb2}
	\end{minipage}
	\caption{Example of ambiguous syntax constructs}
	\label{lst:amb}
\end{figure}
\renewcommand{\figurename}{Fig.}

Due to the above mentioned syntactic ambiguities there are limits towards how similar the bytecode obtained via \jess can be to the original bytecode.
This limit highly depends on how much of the code base is actually available and how much has to be supplemented via stub generation.
\jess was designed to generate bytecode that is identical or at least very close to the original one.
However, since this is not always possible, \jess offers the option to mark the locations where ambiguities may have caused imprecision.
In case of an unknown FQN, \jess sets the package of the generated stub to a special \texttt{unknown} package.
By doing so, the \texttt{unknown} package within the bytecode can be treated as a wildcard in a potential comparison and therefore still be compared against the original bytecode.
Whenever a type (e.g. return or parameter type of a method) cannot be precisely inferred, \jess sets the type within the generated stub to a special \texttt{Unknown} type.
Whenever an ambiguity is detected, \jess can optionally perform this unknown type assignment.
	\newcommand{\RQONE}{How does \jess's compilation perform on popular and current Java projects?}
\newcommand{\RQTWO}{How similar is the bytecode obtained via \jess to the original bytecode?}
\newcommand{\RQTHREE}{To which degree can \jess enable the compilation of commit changes?}

\section{Evaluation}
\label{sec:evaluation}

In the following we evaluate the effectiveness of \jess's compilation.
To do so, we implemented our approach in a tool.
We answer the following research questions.
\begin{enumerate}
	\item[\textbf{RQ1:}] \RQONE
	\item[\textbf{RQ2:}] \RQTWO
	\item[\textbf{RQ3:}] \RQTHREE
\end{enumerate}
The first two research questions focus on evaluating the performance of \jess in general.
They aim at showing how effectively it can be used on partial code bases and how similar the generated bytecode is to the one obtained by a successful compilation of the full project.
The third research question focuses on the use case of commit compilation, to evaluate its effectiveness to compile code changed in commits fixing known vulnerabilities.

\subsection{RQ1: \RQONE}
\label{subsec:rqone}

\begin{figure}
	\centering
	\includegraphics[width=0.8\linewidth]{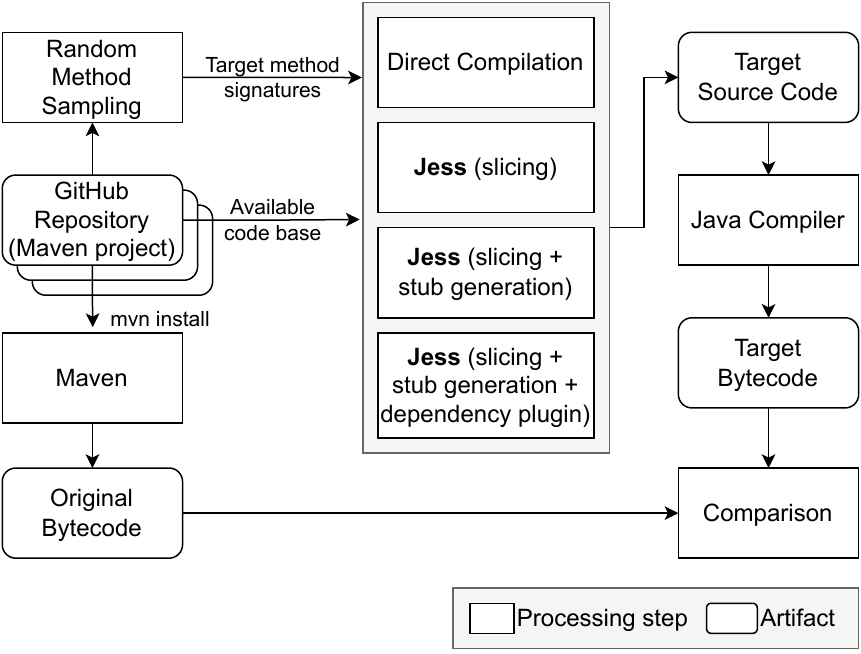}
	\caption{Overview of our evaluation setup for RQ1 and RQ2}
	\label{fig:rq1rq2overview}
\end{figure}

In this research question we aim at investigating how \jess performs on the latest state of Java projects hosted in source code repositories.
Figure~\ref{fig:rq1rq2overview} illustrates our experimental setup for answering RQ1 and RQ2.
Based on stars, we selected the 1,000 most popular Java projects on GitHub.
From these 1,000 projects we selected the Maven projects~\cite{maven}, which resulted in 347 total projects considered in our evaluation.
We cloned each of the repositories at their most current state and, for each repository, we randomly sampled methods as individual target for compilation via \jess.
To have a diverse but manageable dataset, we randomly selected 100 distinct methods or constructors from each repository.
We use ``methods'' to refer to both, methods and constructors, from here on.
To omit trivial methods like getters and setters, we only considered methods containing at least three lines of code in their body.
The code base contained within the cloned repository and the signatures of the sampled target methods are input to \jess.
On the same set of target methods, we performed four different compilation procedures.
To establish a baseline, we directly provided the target file to the Java compiler for compilation, without processing it with \jess first.
We then used \jess to perform three different compilation experiments.
In the first experiment, we only used the Java source code within the cloned repository as available code base and solely performed slicing without generating any stub files.
In the second experiment, we used the same code base, but complemented the slicing with stub generation.
In the final experiment, we added a simple Maven plugin, which (transitively) downloads the available dependencies configured within the project's pom.xml files at the version configured within the project.
Note that the custom plugin does not perform any dependency conflict resolution nor does it guarantee that all configured dependencies are available.
We use this custom plugin as we cannot use the default Maven Dependency Plugin~\cite{mavenDependencyPlugin} for our experiment, as it employs a fail early strategy and stops the dependency resolution process as soon as a single artifact resolution fails.
Then, before running \jess, the downloaded JAR files are added to the available code base.
In the next step, the target source code generated by \jess (i.e., sliced target file, sliced support files, and optional stub files) is handed to the Java compiler for compilation.
To evaluate the similarity of the target bytecode obtained via \jess to the original one (cf. RQ2 in Section~\ref{subsec:rqtwo}), we invoke the default \texttt{mvn install} command on each of the cloned projects to compile it via the provided build scripts wherever possible.
We then compare the target bytecode to the original bytecode.
As previously reported (see Section~\ref{sec:introduction}), automatic compilation via invoking the provided build scripts often results in failure, even though the selected projects are at their most current state.
Thus, we are only able to perform a comparison for the projects where the default build scripts yielded a successful compilation, which are 134 of 347 projects.
We report on the results of the bytecode comparison in Section~\ref{subsec:rqtwo}.
Our experiments were executed inside a Docker container running Alpine Linux with Eclipse Temurin JDK 17 and Maven 3.9.3.
The container was executed on a Debian 10 machine, configured to use four cores of an Intel Xeon E5-2695 v3 (2.30 GHz) CPU and 32GB of main memory.

Table~\ref{tab:rq1results} shows the results of our compilation experiments (RQ1).
In total we sampled 32,970 distinct methods originating from 21,856 different files.
The number of methods listed in Table~\ref{tab:rq1results} is slightly lower than 100 per project, since a few small projects did not contain enough methods with at least three lines of code.
The first row of the table shows our baseline direct compilation.
When providing the file containing the target method directly to the Java compiler the success rate was very low at 6.79\%, since most files contain references to other files.
The compilation success rate increases to 38.43\%  by slicing the file via \jess first.
In fact, references to files outside of the available code base, but not required for compilation, are removed from the respective target and support files.
This result shows that a considerable amount of methods within a project can be compiled solely relying on the source code within its repository, without the need to resolve external dependencies at all.
However, if the target method refereces an external dependency, which is not available within the provided code base, compilation fails without stub generation.
Applying stub generation after slicing, the success rate of \jess significantly increases to 62.30\%.
In our final experiment we run our custom dependency download plugin beforehand, to increase the available code base as much as possible. 
By increasing the code base available to \jess through the downloaded JAR files, the compilation success rate increases to 72.02\%, indicating that \jess allows for a targeted compilation of most methods within the investigated projects.

We investigated the cases in which \jess failed (even when using the dependency plugin) and found the main reasons to be: no longer available dependencies, unresolved dependency conflicts, and limitations to the implementation of \jess itself. 
Our implementation of \jess relies on the symbol-solving capabilities of the popular Javaparser~\cite{javaParser} library, thus it inherits Javaparser's limitations, e.g., lack of support for all up-to-date Java features and problems in resolving language constructs like method references (:: operator), lambda expressions, and variable arguments (Varargs).
These limitations to Javaparser especially apply in partial code bases.

We additionally measured the time that \jess and the subsequent compilation of the target method require (see Table~\ref{tab:rq1results}).
As expected, since no resolution, slicing, or stub generation needs to be performed, a direct compilation of the target file is very fast.
However, the success rate of a direct compilation is low.
Only applying slicing, the average processing time of \jess (including compilation) takes 400ms.
As expected the required time slightly increases to 464ms when \jess additionally applies stub generation.
Furthermore, as the code base that needs to be considered for resolution increases when using the dependency plugin, the required processing and compilation time also increases to 831ms (not including dependency download time).
These results show that even when a compilation of the full project is possible, compilation via \jess may be preferable if the area of interest within the code is limited due to its fast compilation times.

\llbox{\jess was able to compile up to 72\% of randomly sampled methods from popular Java projects, in isolation, only taking 831.3ms per method on average.}

\begin{table}[]
	\centering
	\begin{tabular}{l|ccc}
		& \textbf{Methods} & \textbf{Compilable} & \textbf{\begin{tabular}[c]{@{}c@{}}Avg. Time\\ (ms)\end{tabular}} \\ \hline
		\textbf{Direct Compilation}   & 32,970 & 2,238 (6.79\%)   & 125.7 \\
		\textbf{Jess (slicing)}       & 32,970 & 12,670 (38.43\%) & 400.9 \\
		\textbf{Jess (+ stub gen.)}   & 32,970 & 20,539 (62.30\%) & 464.5 \\
		\textbf{Jess (+ dep. plugin)} & 32,970 & 23,744 (72.02\%) & 831.3
	\end{tabular}
	\caption{Compilation success rates using \jess}
	\label{tab:rq1results}
\end{table}

\subsection{RQ2: \RQTWO}
\label{subsec:rqtwo}

In this research question we investigate the similarity of the bytecode obtained via \jess to the original bytecode, when the source code is compiled the intended way via provided build scripts.
For this investigation we used the same experimental setup as previously described in Section~\ref{subsec:rqone}.
Wherever we were able to successfully compile a target method via \jess and by invoking the provided Maven build scripts, we compared the bytecode of the target method originating from \jess and from the default build.
We used ASM 9.5~\cite{asm} to extract the textual representations of the bytecode corresponding to the respective target methods (based on the signature) and performed a textual head-to-head comparison where the method's bytecodes are considered different as soon as there is a single textual difference between the compared bytecodes.
Note that, since bytecode is a binary format, the extracted textual representation depends on the ASM framework's interpretation, which is based on the Oracle JVM specification~\cite{jvmClassFileFormat}.
To see how similar the resulting bytecodes are, if they are not equal, we computed the normalized Levenshtein Distance (NLD)~\cite{yujian2007normalized} on their textual representations.
The NLD is a commonly used measure that can be used to calculate the similarity of two text sequences.
The Levenshtein Distance represents the number of characters that need to be inserted, deleted or substituted to transform one text sequence into the other.
The normalized Levenshtein Distance in addition also considers the lengths of the text sequences to calculate a percentage value of similarity.
An NLD of 100\% indicates that every single character needs to be changed, while a value of 0\% indicates that both sequences are identical.
The lower the NLD is, the more similar are the compared text sequences.

Table~\ref{tab:rq2results} shows the results of our comparison.
The first row of the table shows the comparison results when only slicing is applied by \jess.
The number of compared methods is smaller than investigated in Section~\ref{subsec:rqone}, as we were only able to compile 134 of the 347 projects using their provided build scripts.
Out of the compared 7,722 methods, 92.75\% yielded the exact same bytecode as obtained through the intended build.
Furthermore, the compared methods were highly similar to each other, as the average NLD is 0.29\%, indicating a very high degree of similarity between the respective method's bytecodes.
The second row of Table~\ref{tab:rq2results} shows the results when stub generation is applied in addition.
As expected, when stubs need to be generated to enable a successful compilation, the degree of similarity decreases through the ambiguities that arise.
Out of the 11,519 compared methods, 76.17\% of the methods yielded identical bytecode.
The average NLD increased to 1.85\%.
Though higher than in the case of pure slicing, the bytecode obtained via \jess is still very similar to the original bytecode, while increasing the number of compilable methods by more than 62\%, indicating the added value of stub generation.
Finally, the last row of the table shows the comparison results when the code base available to \jess is augmented with resolvable external dependencies.
Out of the 13,630\footnote{One source code method may expand into multiple bytecode methods after compilation, e.g. due to anonymous classes} compared methods, 12,192 (89.45\%) yielded the exact same bytecode.
Furthermore, the average NLD decreased to 0.66\%.
As expected, due to the availability of external JAR files, the similarity increases compared to the experiment with slicing + stub generation as certain references become resolvable and thus fewer stubs are required for compilation.

We investigated the methods that differ from their original bytecode after purely applying slicing via \jess.
While it is clear that, due to ambiguities, the bytecode obtained through compilation supplemented with stubs may contain differences, it is not obvious when only slicing is applied.
Our investigation revealed the main causes of the remaining differences to be:
\begin{itemize}
	\item Properties that are reflected into the bytecode when compiling using Maven~\cite{mavenResourcesPlugin}.
	\item Bytecode that is automatically generated by frameworks (e.g. non-null checks), which are configured in the provided build scripts and invoked by Maven, e.g. Project Lombok~\cite{lombok} or Checker Framework~\cite{checkerFramework}.
	\item Anonymous classes and synthetic methods generated by the Java compiler are named with an incrementing numeric suffix (e.g. \texttt{id\$0}).
	This numeric suffix is based on the location within the source code file.
	Since \jess's slicing removes some of these constructs if not required for compilation, the naming of these constructs may differ.
	However, this usually only results in a single digit that differs from the original bytecode, making the bytecode still highly similar to the original one.
\end{itemize}

\llbox{
	Depending on the extent of the available code base, the bytecode obtained via \jess on average only differs by 0.29--1.85\% from the original bytecode with 76--93\% of all compared methods even yielding identical bytecode.
}

\begin{table}[]
	\centering
	\begin{tabular}{l|ccc}
		& \textbf{Methods} & \textbf{Equal} & \textbf{Avg. NLD} \\ \hline
		\textbf{Jess (slicing)}       & 7,722  & 7,162 (92.75\%)  & 0.29\% \\
		\textbf{Jess (+ stub gen.)}   & 11,519 & 8,774 (76.17\%)  & 1.85\% \\
		\textbf{Jess (+ dep. plugin)} & 13,630 & 12,192 (89.45\%) & 0.66\%
	\end{tabular}
	\caption{Similarity of the bytecode obtained via \jess to the original bytecode}
	\label{tab:rq2results}
\end{table}
\subsection{RQ3: \RQTHREE}
\label{subsec:rqthree}

\begin{figure}
	\centering
	\includegraphics[width=0.8\linewidth]{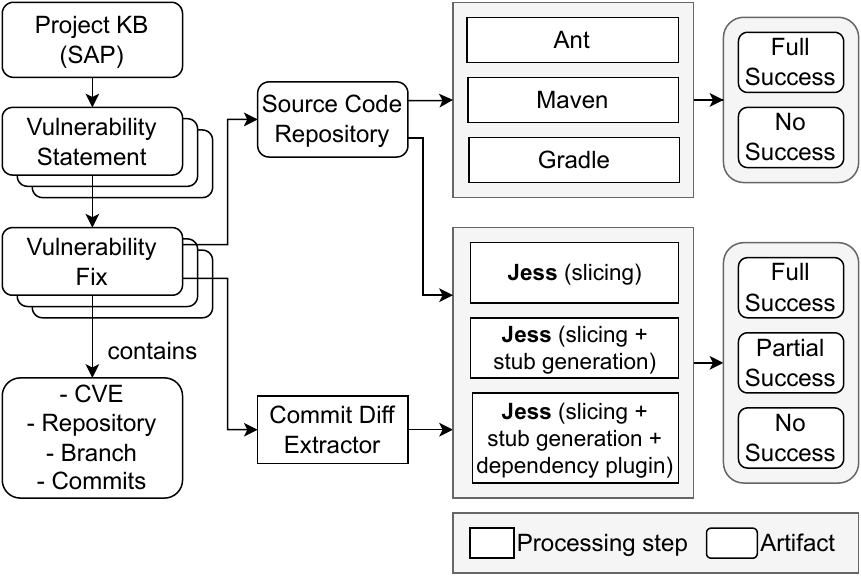}
	\caption{Overview of our evaluation setup for RQ3}
	\label{fig:rq3overview}
\end{figure}

In this research question we evaluate the primary use case \jess has been designed for: to compile commit changes within repository snapshots.
Figure~\ref{fig:rq3overview} shows an overview of our experimental setup.
We considered SAP's Project KB~\cite{ponta2019manually} as dataset for our evaluation.
Project KB is a manually curated database, which contains vulnerabilities affecting industry-relevant open source projects from the years 2005--2023.
In total, it contains 1,297 \textit{vulnerability statements}, which map a Common Vulnerability Enumeration (CVE)~\cite{nvd} to the commits within a project's source code repository that fix the corresponding vulnerability.
Such commits are referred to as \textit{fix-commits}.
As described previously, Project KB only contains the snapshot version with the patch applied, but not the corresponding release version.
In this experiment we thus assess \jess's performance on directly compiling these snapshots.
Each commit in Project KB changes classes and methods that are related for some security vulnerability, making all of them compilation targets for vulnerability detection tasks.
In our evaluation, we only consider statements that affect Java projects and contain corresponding fix-commits, which leaves us with 693 unique vulnerability statements.
Each vulnerability statement consists of one or more \textit{vulnerability fixes}, each containing a set of fix-commits that fix a certain CVE.
In our case, the 693 considered statements consist of 920 vulnerability fixes.
We cloned the source code repository belonging to each vulnerability fix, checked out the most recent commit out of the set of fix-commits, which we continue to refer to as \textit{snapshot}, and applied one of two subsequent processing steps to the snapshot.
At first we are interested in establishing a baseline of how many of these snapshots are able to be compiled by using the build scripts provided within each project.
We used the same procedure as Hassan et al.~\cite{hassan2017automatic} to automatically locate the appropriate build scripts and invoke them in a systematic way.
We considered the three popular Java build tools with the respective build commands that provided the highest chance of success as reported by Hassan et al.: Maven~\cite{maven} (\texttt{mvn compile}), Gradle~\cite{gradle} (\texttt{./gradlew} and \texttt{gradle build}) and Ant~\cite{ant} (\texttt{ant build}).
Moreover, since Ant does not define a default build command like Maven and Gradle do, we added the command \texttt{ant compile} to our evaluation after manually inspecting the respective build scripts to boost the compilation success rate.
In total 907 of the 920 repositories within our dataset contain build scripts and are thus theoretically compilable via invoking build scripts.
Further, we provided the same repository snapshots to \jess.
Additionally, we performed a \textit{commit diff extraction} in which we extracted the signatures of classes, methods, constructors and fields that have been modified or added within the respective set of fix-commits.
We then separately handed each file that has been modified within the fix-commits to \jess as target file.
Furthermore, we specified all modified class members, as determined by the commit diff extraction, as target members.
Like in our previous experiments, we executed \jess with three different configurations (see Section~\ref{subsec:rqone}) and provided the repository as available code base.
The outcome of invoking a build script is typically binary, either the compilation is successful and one obtains the bytecode of all files within the snapshot or the compilation is not successful and one obtains no bytecode at all.
However, when compiling commit changes via \jess the outcome is more fine-grained.
When all files that have been modified within the set of fix-commits were successfully compiled, we consider it as a full success.
If no file could be successfully compiled we consider it as no success.
However, in case \jess was able to compile a subset of the targets, we consider it a partial success.

Table~\ref{tab:rq3buildTools} shows the results when invoking the provided build scripts for compilation.
The results are split up for each individual build tool and show how many snapshots could successfully be compiled by invoking the provided build scripts and how many of the files, modified in the fix-commits, could thus be compiled.
One can immediately see that the rate of successful compilations is rather low, with projects using Ant as build tool having by far the highest success rate and project's using Gradle having the lowest.
An investigation of the success rate for Ant projects revealed that almost all of the successful compilations (27) can be attributed to snapshots of the Apache Tomcat~\cite{tomcat} repository, which is exceptionally well maintained.
In total, fewer than 15\% of the snapshots could be compiled using the provided build scripts, only yielding the bytecode of 8.8\% of files changed in the commits, which is a noticeably lower success rate than reported in other studies~\cite{sulir2016quantitative, hassan2017automatic, tufano2017there}.
However, Project KB contains entries dating back to 2005, which means that many of the fix-commits are old.
According to Sul{\'\i}r and Porub{\"a}n~\cite{sulir2016quantitative} the age of the project heavily influences the probability of successful compilation, which could explain the low success rates.

\begin{table}[]
	\centering
	\begin{tabular}{lcc}
		\textbf{Build Tool} & \textbf{Compilation Succ.} 	& \textbf{Compiled Files} 	\\ \hline
		Ant    				& 29 / 84 (34.5\%)   			& 55 / 239 (23.0\%)  		\\
		Maven  				& 94 / 657 (14.3\%)  			& 174 / 2,133 (8.2\%) 		\\
		Gradle 				& 12 / 166 (7.2\%)   			& 27 / 542 (5.0\%)  		\\ \hline
							& 135 / 907 (14.9\%) 			& 256 / 2,914 (8.8\%) 
	\end{tabular}
	\caption{Compilation success rates when invoking build scripts provided by the projects}
	\label{tab:rq3buildTools}
\end{table}

Table~\ref{tab:rq3jess} shows the compilation results using \jess.
Again, each row represents the different configurations used for \jess.
One can immediately see that \jess performs significantly better than the provided build scripts.
Even when only slicing is applied, \jess was able to fully compile all commit changes within 23.6\% of snapshots.
For 53\% of snapshots it was able to compile it at minimum partially, resulting in at least some successfully compiled files.
When comparing the amount of files \jess was able to compile to the number of files compilable via provided build scripts there is an even higher difference.
This difference only increase further when \jess additionally applies stub generation or uses the dependency plugin to download available JAR files.
Note that the dependency plugin only works for Maven projects, so for the Ant and Gradle projects no dependencies are downloaded at all.
When using \jess with slicing, stub generation and the custom dependency plugin, it is able to fully compile all commit changes within 53\% of the snapshots, while being at least able to compile parts of 80.4\% of all snapshots.
A partial compilation of the fix would allow for a subsequent matching for at least part of the patch.
In total \jess is able to compile 73.6\% of all files modified within the fix-commits for Java projects listed within Project KB.
Note that due to only being able to obtain the bytecode of at most 174 files modified within fix-commits (less than 6\% of files), using \jess \textit{and} the provided build scripts, we did not perform another bytecode comparison experiment.
The results from the previous experiment (see Section~\ref{subsec:rqtwo}) should also apply here.

In theory, the success rate of \jess can be increased even further.
As long as all constraints imposed by the compiler are satisfied, the compilation will succeed, which can be achieved by e.g. altering the source code designated for compilation.
However, as constraints are not unambiguous in partial code bases (see Section~\ref{subsec:ambiguities}), this always comes with a trade-off between success rate and degree of similarity of the resulting bytecode to the original one. 
As \jess was primarily designed for producing bytecode that is suited for a subsequent comparison, we privilege bytecode similarity over success rate.
\jess is able to compile 73.6\% of fix-commits contained in Project KB, with the resulting bytecode having a high degree of similarity to the original one (see Section~\ref{subsec:rqtwo}).

\llbox{
	Relying on the provided build scripts only allows for a compilation of 8.8\% of files affected by fix-commits within Project KB.
	Using \jess we were able to obtain the bytecode of the fixes within 73.6\% of all files.
	For 80.4\% of vulnerability fixes \jess was able to compile at least parts of the fix, allowing for a subsequent matching of the patch.
}

\begin{table}[]
	\centering
	\resizebox{\linewidth}{!}{%
	\begin{tabular}{lccc}
		\multicolumn{1}{l|}{\textbf{}}        	& \textbf{Full Succ.} 	& \textbf{Partial Succ.} 	& \textbf{Compiled Files} \\ \hline
		\multicolumn{1}{l|}{\textbf{Slicing}} 	& 214 / 907 (23.6\%)  	& 481 / 907 (53.0\%)     	& 1,258 / 2,914 (43.2\%)  \\
		\multicolumn{1}{l|}{\textbf{+ SG}} 		& 412 / 907 (45.4\%) 	& 705 / 907 (77.7\%) 		& 1,939 / 2,914 (66.5\%) \\
		\multicolumn{1}{l|}{\textbf{+ DP}}		& 481 / 907 (53.0\%) 	& 729 / 907 (80.4\%) 		& 2,146 / 2,914 (73.6\%) \\
		\multicolumn{4}{c}{SG = Stub Generation, DP = Dependency Plugin}                                     
	\end{tabular}
	}
	\caption{Compilation success rates via \jess}
	\label{tab:rq3jess}
\end{table}

	\section{Related Work}
\label{sec:relatedWork}

Different approaches have been developed, which aim at increasing the rate of successful compilations of Java projects by fixing the default build process or enabling a compilation of partial code bases.
Stubber~\cite{schafer2021stubber} aims at compiling partial source code bases via providing predefined stub classes.
In contrast to \jess, Stubber does not specifically adjust the stub classes to the target file, but modifies the existing source code to fit a previously created stub class, thus changing the resulting bytecode of the target file.
Jcoffee~\cite{gupta2020jcoffee} leverages the error messages reported by the Java compiler during compilation.
Based on the received error, Jcoffee modifies the source code or generates needed declaration code to supplement the compilation.
Stubber and Jcoffee modify the source code of the compilation target within the partial program and therefore alter the resulting bytecode.
Dagenais and Hendren~\cite{dagenais2008enabling} propose a technique that enables static analysis of partial Java programs by applying a similar type inference approach as \jess does during its stub generation step.
Their approach supports up to Java 5 compliant source code, parses the program into a Jimple based abstract syntax tree and adds missing type information directly to the resulting AST.
These approaches use similar strategies as \jess, however, they do not aim at producing bytecode identical to the original compilation.
Thus, these approaches are not applicable to the use case \jess has been designed for.
Melo et al.~\cite{melo2020type} propose a similar concept as Dagenais and Hendren for inferring types of only partially available C programs.
SnR~\cite{dong2022snr} also applies type inference strategies to compile incomplete Java code snippets.
However, in contrast to other techniques that generate supplemental stub files, SnR leverages a pre-built knowledge base, which contains the type signatures of various libraries.
Hassan and Wang~\cite{hassan2017mining} leverage natural language processing techniques to mine readme files and Wiki pages of software repositories to extract build commands that deviate from default build commands.
By doing so they are able to increase the build success rate of projects where all build errors can be solved by running a different set of build commands.
HireBuild~\cite{hassan2018hirebuild} uses a history-driven repair approach to fix failing Gradle build scripts.
It uses revisions from a build fix database that have previously been applied to build scripts in order to fix them and automatically creates new build fix templates from the successful fixes.
These templates are then automatically applied to the failing build script if a similar failure mode as in the past is causing the build failure.
Lou et al.~\cite{lou2019history} investigated HireBuild and found multiple shortcomings.
Inspired by the findings they propose a new technique called HoBuFF that does not rely on historical information but generates patches based on resources available within the present project.
BuildMedic~\cite{macho2018automatically} tries to fix dependency-related build issues in Maven projects by implementing an approach that automatically updates or deletes included dependencies.

Hassan et al.~\cite{hassan2017automatic} investigate the feasibility and challenges that arise for the automatic building of Java projects in software repositories.
Their study shows that automatic building is a challenging task, which often results in failure.
Tufano et al.~\cite{tufano2017there} specifically investigate the compilability of snapshots within Java software repositories.
Their study shows that almost all of the investigated projects contain long periods of snapshots that are not compilable using the provided build scripts.
Sul{\'\i}r and Porub{\"a}n~\cite{sulir2016quantitative} study automatic build invocation on more than 7,000 Java projects.
Their investigation reveals that projects using Gradle or Ant as build tools have higher build failure rates than projects using Maven.
Furthermore, they show that dependency resolution issues are by far the most common issue, which is causing build failures.
	\section{Threats to Validity}
\label{sec:threats}

Our evaluation of \jess's performance on compiling commit changes relies on the vulnerability database Project KB.
Project KB contains fix-commits affecting industry relevant open source projects.
This means that the projects used within our evaluation are typically well maintained. 
An evaluation on projects that are not industry-relevant, may have yielded different results.
Furthermore, we only used default build commands (except for Ant) for invoking builds, as Hassan et al.~\cite{hassan2017automatic} reported them to have the highest success rate for compilation.
However, in our specific dataset there may still be different build commands, which could have caused a higher compilation success rate when invoking the provided build scripts.
Moreover, we only used Maven projects for our first two evaluation experiments, as we only created a simple dependency download plugin for Maven.
While such a plugin can also be created for Gradle and Ant, the performance of dependency resolution may be different for other build tools.
	\section{Conclusion}
\label{sec:conclusion}

In this paper we presented our approach \jess, which enables the compilation of commit changes within Java source code repositories.
By resolving all references made from the source code parts changed in the commit, \jess slices away the parts of the code base which are not needed for a successful compilation.
If references are not available in the provided code base, \jess can generate stub files which contain type stubs that are inferred from their usage context.
By combining these steps, our approach can compile exactly the areas of interest which have been changed in a commit.

We evaluated \jess on 347 of the most popular Java GitHub projects.
Of the 32,970 randomly sampled methods and constructors, \jess was able to compile more than 72\% of which the resulting bytecode on average only differed by 0.66\% from the bytecode obtained when running the build scripts provided in the code bases. 
In almost 90\% of cases \jess produced equal bytecode.
In a real-world experiment using the Project KB vulnerability database, the provided build scripts only achieved successful compilation of 8\% of all files modified within the provided fix-commits. 
\jess allowed for the compilation of 73\% of all files without the need to invoke any build scripts.
These results show that \jess can effectively compile commit changes of which the resulting bytecode closely resembles the original one, whereas the provided build scripts often do not allow for a compilation at all.
	
	\section*{Acknowledgment}
	This work was partially supported by the German Research Foundation (DFG) within the Collaborative Research Centre ''On-The-Fly Computing`` (GZ: SFB 901/3) under the project number 160364472.
	
	\bibliographystyle{plainurl}
	\bibliography{literature}
\end{document}